\documentclass[aps,pre,onecolumn,showpacs,a4paper]{revtex4}
\usepackage{epsfig}
\usepackage{graphicx}

\newcommand{\thickone}{\mbox{$1\!\!1$}}
\newcommand{\square}[8]{
\setlength{\unitlength}{1.01cm}
\begin{picture}(5,3.6)
\thicklines

\put(0,3){\makebox(0,0){$#1$}}
\put(5,3){\makebox(0,0){$#2$}}
\put(0,0){\makebox(0,0){$#3$}}
\put(5,0){\makebox(0,0){$#4$}}

\put(-.5,1.5){\makebox(0,0)[r]{$#6$}}
\put(5.5,1.5){\makebox(0,0)[l]{$#7$}}
\put(2.5,-0.5){\makebox(0,0)[t]{$#8$}}
\put(2.5,3.5){\makebox(0,0)[b]{$#5$}}

\put(1,0){\vector(1,0){3}}
\put(1,3){\vector(1,0){3}}
\put(0,2.5){\vector(0,-1){2}}
\put(5,2.5){\vector(0,-1){2}}
\end{picture}
}

\begin{document}

\title{Transfer matrices for the totally asymmetric simple exclusion process}

\author{Marko Woelki\dag\ddag\footnote[1]{woelki@lusi.uni-sb.de} and Kirone Mallick\dag}

\affiliation
{\dag Institut de physique th\'eorique, Centre d'\'Etudes Atomiques,
 F-91191 Gif-sur-Yvette, France}
\affiliation
{\ddag Fachrichtung Theoretische Physik, Universit\"at des Saarlandes, 66123 Saarbr\"ucken, Germany}


\begin{abstract}
 We consider  the totally asymmetric simple exclusion process (TASEP)
 on a finite lattice  with open  boundaries. We show, 
 using the  recursive structure of the  Markov matrix  that encodes the dynamics, 
  that there  exist two transfer matrices  $T_{L-1,L}$  and
 $\tilde{T}_{L-1,L}$  that  intertwine the Markov matrices of  consecutive system sizes: 
 $\tilde{T}_{L-1,L}M_{L-1}=M_{L}T_{L-1,L}$. 
  This semi-conjugation  property  of the dynamics provides an  algebraic counterpart 
  for  the matrix-product representation
  of the steady state of the process. 

\end{abstract}

\maketitle

\section{Introduction}

 Non-equilibrium statistical mechanics has progressed a lot thanks to  the 
 study  of interacting particle processes in one dimension \cite{spohn,liggett,PaulK}.
  In this field,   the  asymmetric simple exclusion process (ASEP) has been  playing  
 a paradigmatic role with  an    impressive body of knowledge 
  accumulated  during the last twenty years 
  \cite{zia,derrida,EvansBraz,SchutzRev}. This model can be  investigated 
  from  very  different points of view: 
  simple exclusion  indeed  has a   complex story \cite{varadhan}. 
 Two  noteworthy  techniques  are  the Bethe Ansatz \cite{Dhar,GwaSpohn}, useful  for 
  spectral properties of the dynamics, 
 and the matrix  Ansatz, initiated in \cite{DEHP}, in which  
  the stationary measure of the model
 is represented as a matrix-product state. The  matrix 
  Ansatz first   appeared  in the study of   the totally
   asymmetric simple exclusion process (TASEP) on a finite 
  discrete lattice  with open  boundaries,  
 as a trick to represent  the  stationary weights  of the configurations.  
 It was  observed, empirically, 
 \cite{DDM,Schutz} that the weights of configurations of sizes $L$ and $L-1$ are
 related through  recursion relations that can be suitably encoded in  a matrix-product
 state. Subsequently, this  idea  bloomed 
 into a fruitful and powerful technique that can be summarized as follows:
 given a stochastic model,  look  for a  suitable algebra  to 
 represent its steady state.   A recent and exhaustive review
 of the matrix-product representation for stochastic 
 nonequilibrium systems can be found in  \cite{Blythe}.

 However, the very fact that the weights of  configurations of 
 {\it different}  sizes can be  related through  some combinatorial identities 
 is deeply  puzzling;  indeed,   two models of different sizes have a priori 
 no relation at all  with  one another:  their phase  spaces are
  totally disconnected (the dynamics conserves the  size of the system).  
 Moreover, the matrix  Ansatz does not seem to be logically related to the structure of
 the Markov operator: the algebra  used to represent the  stationary weights
  has  often to be determined by inspection of 
 simple cases   or by analogy with   known examples
   \cite{Derrida2class,degier,MMR,chikashi1,woelkirand,woelkipar} (the review   \cite{Blythe} that 
 describes in detail most of the  exact solutions in the field 
  is  particularly helpful in this respect). 
  Once the    algebra is found,   the steady state is written as   a trace
  over this algebra and  is shown to   vanish  under the action of the  Markov matrix.
  This  last step of the proof  involves a  cancellation mechanism
  usually requiring  an auxiliary ('hat')
  algebra \cite{hinrichsen,nikolaus}. It has  been  shown rigorously
   that  most of the models  admit  formally  a  matrix  Ansatz  but,  unfortunately,
  the proof is not constructive \cite{krebs}.

  The aim of the present work is to show  for the TASEP with open boundaries that 
 the Markov matrices of  two  consecutive system sizes 
  are semi-conjugate of  each other through  two transfer
 matrices.  This conjugation property is a characteristic  of the dynamics  and   
 it only relies on the recursive structure  of the  Markov matrix. 
 A similar property has been proved 
 for the multi-species exclusion process on a ring for
  which the matrix-product representation  involves complicated
 tensor products of quadratic algebras \cite{pablo,sylvain,chikashi}. Recently, 
  a  conjugation property has been used   to derive exact results for an 
 annihilation model for which a   matrix  Ansatz could
  not  be found \cite{arvind}.  We believe
 that the existence of a  dynamical conjugation  that relates a given 
 model to a simpler one (simpler because it involves  a smaller number of sites, or 
 of particles,  or of types  of particles...) 
 is a   fundamental property
 that underlies  the solvability of many nonequilibrium processes.

\section{Recursive structure of the TASEP  Markov matrix}

 We first recall the dynamical rules of the 
 TASEP in a  one-dimensional discrete lattice of size $L$ with open boundaries. 
 In this model,   particles are injected at  site
1 at rate $\alpha$ and rejected from site $L$ at rate $\beta$. Every
site can be occupied by at most one particle. Particles in the bulk can hop  
stochastically  with rate 1 from a  site to 
 the adjacent  site on its right if it  is vacant.
 The phase space $\Omega_L$ of the system consists of $2^L$ different configurations.

The Markov matrix for a general stochastic system on a 1d lattice of size
$L$ with open boundaries and nearest-neighbor interactions is given by \cite{krebs}
\begin{equation}
M_L = h_L^{\rm left} + \sum\limits_{i=1}^{L-1} h_L(i,i+1) + h_L^{\rm
right},
\end{equation}
with
\begin{eqnarray}
h_L^{\rm left}&=&h^{(l)} \otimes \thickone_{L-1}, \quad
h_L^{\rm right}=\thickone_{L-1} \otimes h^{(r)},\\
h_L(i,i+1)&=& \thickone_{i-1}\otimes h \otimes \thickone_{L-i-1}.
\end{eqnarray}
 Here, we  restrict ourselves to a two-dimensional state space. Thus $h^{(l)}$ and
$h^{(r)}$ are $2\times 2$ matrices reflecting the boundary
interactions and $h$ is the local $4\times 4$ matrix for the bulk.
More precisely, we have:
\begin{eqnarray}
h^{(l)} =  \left( \begin{array} {ll} -\alpha & 0 \\ \,\,\,\, \alpha & 0 \end{array} \right), 
 \quad h^{(r)} =  \left( \begin{array} {ll} 0  & \,\,\,\, \beta \\  0
  & -\beta \end{array} \right)
 \quad \hbox{and} \quad h  =  \left( \begin{array} {llll}
                                   0 & 0 &  \,\,\,\, 0 & 0 \\
                                   0 & 0 &   \,\,\,\,1 & 0 \\    
                                    0 & 0 & -1 & 0 \\
                                    0 & 0 & \,\,\,\, 0 & 0    \end{array} \right) \, .
 \label{localmat}
\end{eqnarray}
Throughout this paper we use $\thickone_d$ for the $2^d$ dimensional
identity but for $d=1$ the index is supressed. We observe  that
 the following  recursion for the Markov matrix is satisfied 
\begin{equation}
\label{rec} M_L=\thickone \otimes M_{L-1} - \thickone \otimes
h_{L-1}^{\rm left} + h_{L-1}^{\rm left} \otimes \thickone +
h_{L-1}(1,2)\otimes \thickone.
\end{equation}
 We now explain this  formula: a system of size $L$ can be obtained
 by adding  a site with index 0
to the system  of size  $L-1$ between the left reservoir and site 1. Then,
 the Markov matrix $M_L$ for the larger system can be expressed  in terms of 
  $M_{L-1}$. Naively, one would write  $M_L= \thickone  \otimes
M_{L-1}$: this is correct  in the bulk but not in the vicinity of  the left boundary.
 Therefore    this incorrect   formula has to   be repaired as follows: 
(i) particles  now enter  at site 0
 so   we  must subtract the matrix elements that correspond to injecting 
  particles at  site 1 which no longer occurs (second term on the rhs).
  (ii) We add a matrix which makes 
  the particles enter at  site 0 (third term).  (iii)  Finally, 
 particle hopping  from site 0 to site 1 is implemented by the fourth
 term (which  is equal  to  $h\otimes \thickone_{L-2}$).

For the TASEP, substituting the explicit expressions~(\ref{localmat})
 in equation~(\ref{rec}), we are led to 
  the following recursive structure of the Markov matrix:
\begin{eqnarray}
\label{M}
M_L = \left( \begin{array} {cc} M_{L-1}-\nu_{L-1} & \rho_{L-1}\\
\alpha \thickone_{L-1} & M_{L-1} + \omega_{L-1}\end{array} \right),
\end{eqnarray}
where we have  defined
\begin{eqnarray}
\nu_L= \left( \begin{array} {ll} 0 & 0\\
\alpha \thickone_{L-1} & \alpha \thickone_{L-1}\end{array} \right),
\quad \rho_L= \left( \begin{array} {ll} 0 & 0\\
\thickone_{L-1} & 0\end{array} \right)
\quad \rm{and}  \quad  \omega_L=
 \left( \begin{array} {ll} (\alpha-1)\thickone_{L-1} & 0\\
-\alpha \thickone_{L-1} & 0\end{array} \right).
\label{def:nu-rho-omega}
\end{eqnarray}

\section{Transfer matrices for the TASEP}

We shall prove that the  Markov matrices $M_{L}$ and  $M_{L+1}$  for 
 two consecutive system sizes are related by the relation
\begin{equation}
\label{rel} \tilde{T}_{L,L+1}M_{L}=M_{L+1}T_{L,L+1}.
\end{equation}
 This semi-conjugation relation  can be illustrated by the following commutative diagram:
\vglue .55cm
\begin{center}
\square{ \Omega_L   \hskip 1cm}{\hskip 1cm \Omega_L   } 
       { \Omega_{L+1} \hskip 1cm}{\hskip 1cm  \Omega_{L+1}}
       { M_L}{ T_{L,L+1} \hskip -1mm} 
       { \hskip -1mm \tilde{T}_{L,L+1}}{ M_{L+1}}
\label{DiagComm}
\end{center}
\vglue .8cm
\noindent
 We recall that  $\Omega_L$ and  $\Omega_{L+1}$ are
  the phase spaces  for the systems of size $L$ and
 $L+1$,  respectively. The fact that the diagram is commutative
  means that there are two equivalent
 paths  to evolve from a  state $\mathcal{C}_L$ in  $\Omega_L$:
  either one first   applies  the dynamics $M_L$
 and then identifies the result to a state in  $\Omega_{L+1}$  via $\tilde{T}_{L,L+1},$ or one
 imbeds  $\mathcal{C}_L$  in  $\Omega_{L+1}$ via   $T_{L,L+1}$
  and then applies the dynamics $M_{L+1}$ in the larger
 phase space. It is important to note that the two transfer matrices $T$ and $\tilde{T}$
 are different. If they were equal then a whole fraction of  the spectrum of
  $M_L$ would be contained in  $M_{L+1}$  \cite{chikashi} and this  is not true for the TASEP
 with open boundaries as can be verified with  simple  explicit cases:
 the only common eigenvalue shared
 by  $M_L$  and  $M_{L+1}$  is 0, corresponding to the steady state.
 The fact that the two transfer matrices
 are different  is in constrast with   the annihilation model 
  studied in \cite{arvind}  where   $T=\tilde{T}$. 

 Nevertheless, the very existence of this commutative diagram
  expresses the fact that the dynamics for
 sizes $L$ and $L+1$ are semi-conjugate to  each other. 
This property also allows us to construct
 recursively the steady state of a system of size $L+1$ knowing 
that of the system of size $L$. 
 If  $M_L|v_L\rangle=0$,  the vector $|v_{L+1}\rangle$ defined as 
\begin{equation}
|v_{L+1}\rangle = T_{L,L+1}|v_{L}\rangle \, , 
\end{equation}
 satisfies, using  (\ref{rel}), 
\begin{equation}
M_{L+1}|v_{L+1}\rangle = 0 \, . 
\end{equation}
Hence, if  $|v_{L+1}\rangle$ is not the null-vector, it is
 (but for  a normalisation factor) the stationary
 state of  $M_{L+1}$. This justifies the name `transfer matrix' for  $T_{L,L+1}$. We also 
 remark that $\tilde{T}_{L,L+1}$ has played no role 
in this construction. One could use $\tilde{T}$ 
 as a transfer matrix for the left ground state, but for 
the  Markov matrix  it is known that the  left ground state
 is the line-vector with all components equal to 1. 

 We finally emphasize that the commutative diagram that encodes the
  semi-conjugation property  is an intrinsic characteristic  of the
 dynamics and the knowledge of 
 the steady state of the system is not required.  The transfer matrices can be found for
  small systems   by solving a linear system and their existence relies
  on the recursive structure of the Markov  matrix itself.

\section{Proof of the Semi-Conjugation Property}

In this section  we  prove the relation~(\ref{rel})
  by constructing explicitely  the transfer matrices $T$ and $\tilde{T}$. 
 The transfer matrix $T_{L,L+1}$ for $L\geq 1$ is given by
\begin{eqnarray}
\label{T}
 T_{L,L+1}=\left(\begin{array} {l}  \frac{1}{\alpha} \thickone_{L}\\
T_{L,L+1}^{(2)}\end{array} \right) \quad {\rm with } \quad 
T_{L,L+1}^{(2)}=\left(\begin{array} {ll} \thickone_{L-1} &
\thickone_{L-1}\\ 0 & T_{L-1,L}^{(2)}\end{array}
\right)   \quad  {\rm and } \quad  T_{01}^{(2)}:=\frac{1}{\beta}.
\end{eqnarray}
 The matrix  $T_{L,L+1}$ has been constructed to mimic the recursive algorithm
 provided by the quadratic  algebra found in  \cite{DEHP}. 
We also give an expression for $\tilde{T}_{L,L+1}$ in terms of
 an unknown  square matrix $R_{L-1}$:
\begin{eqnarray}
\label{tT}\tilde{T}_{L,L+1}=\left(\begin{array} {l}  \frac{1}{\alpha} \thickone_{L}\\
\tilde{T}_{L,L+1}^{(2)}\end{array} \right)\quad {\rm with}\;
\tilde{T}_{L,L+1}^{(2)}=\left(\begin{array} {ll} \thickone_{L-1} &
\thickone_{L-1}\\ \alpha R_{L-1} & -R_{L-1}(M_{L-1}-\nu_{L-1})\end{array}
\right) \, .
\end{eqnarray}
  Using   these   expressions of $T$ and $\tilde{T}$, we  calculate
 the left hand side and the right hand side of   equation~(\ref{rel}).
  Both sides are  block rectangular 
 matrices of size 4 by 2 with  elements given  in terms  of the matrices at level $L-1$.
   In order to satisfy   equation~(\ref{rel}), the 8 elements of the l.h.s. matrix  must be
 equal to the  8 elements of the r.h.s. matrix. Amongst the  8 conditions thus obtained,
 6 are tautologically true. The 7th relation is given by
 \begin{eqnarray}
\label{eq:relat1}
\rho_{L-1}T_{L-1,L}^{(2)} = \frac{\nu_{L-1}}{\alpha} \, .
\end{eqnarray}
  This equation  is easily verified by induction  by using 
 the explicit expressions~(\ref{def:nu-rho-omega}) of $\rho_{L-1},\nu_{L-1}$
  and  that of   $T_{L-1,L}^{(2)}$ given in    equation~(\ref{T}).
The 8th relation to be satisfied is given by 
 \begin{eqnarray}
\label{eq:relat2}
 \thickone_{L-1} + T_{L-1,L}^{(2)}\left(M_{L-1}+\omega_{L-1}\right) 
= R_{L-1}\left[ \alpha\rho_{L-1}-\left(M_{L-1}-\nu_{L-1}\right)
 \left(M_{L-1}+\omega_{L-1}\right) \right].
\end{eqnarray}
 This relation can be interpreted as a definition of the   unknown  matrix $R_{L-1}$. 
 In order words, the semi-conjugation property~(\ref{rel}) will be proved, if we are
 able to construct a family of matrices $R$ that satisfy equation~(\ref{eq:relat2})
  for each system size. We can rewrite the generic   equation for the unknown $R_L$ as
\begin{equation}
\label{GRH}
G_{L}=R_{L}H_{L},\\
\end{equation}
where we have defined
\begin{eqnarray}
 G_{L}=  \thickone_{L} + \left(M_{L}+\omega_{L}\right)T_{L,L+1}^{(2)} \quad {\rm and} \quad
\label{HL}
H_{L}=\alpha\rho_{L}-\left(M_{L}-\nu_{L}\right)\left(M_{L}+\omega_{L}\right)
 = \nu_{L} M_{L} - M_{L} \left(M_{L}+\omega_{L}\right) \, ,
\end{eqnarray}
 the last equality resulting the fact that  $0 = \alpha\rho_{L} +\nu_{L}\omega_{L}$
 as seen  from equation~(\ref{def:nu-rho-omega}).

 Using  the finer  structure~(\ref{M})  of the Markov matrix and the expression
 for $T_{L,L+1}^{(2)}$ given in equation~(\ref{T}), 
we deduce  the following  recursion for $G$:
\begin{eqnarray}
G_{L+1}=\left(\begin{array} {ll}
 M_{L}-\nu_L+\alpha \thickone_L & M_L + \omega_L+\rho_L\\
 0 & G_L\end{array}\right), 
\label{recursGL}
\end{eqnarray}
which can be iterated to obtain a well defined upper triangular matrix for $G$.
 Therefore all the $G_L$'s are known.  It remains to extract $R_{L}$ from
 equation~(\ref{GRH}).  It is important to note that there is no need
 to calculate $R$ explicitly, since $\tilde{T}$  plays no  role in constructing
 the stationary state 
 (which is entirely  determined  by the right transfer matrix $T$):  the only thing  to
 prove is that   $R_L$ exists, i.e. that the equation $G_{L}=R_{L}H_{L}$
 has at least one solution.
 Thus   equation~(\ref{GRH}) must  satisfy  a solvability 
 condition which here amounts to the fact that any vector in the (right)  kernel
 of $H_{L}$ must  also belong  to the (right) kernel of  $G_{L}$: i.e., 
 any ket $|h_L\rangle$ such that  $H_{L}|h_L\rangle = 0$  must satisfy 
  $G_{L}|h_L\rangle = 0$.  By studying explicitely systems of small sizes,  we found that
 the matrix $H_{L}$ is not invertible and that its kernel consists only
 of   a one-dimensional vector-space.
 We also checked that this  one-dimensional vector-space is included in the kernel
 of $G_{L}$ (which is of dimension $2^L -L -1$).
  Formal proofs of these facts are given in the  appendix.  This  shows that  $R_{L}$  exists
  and concludes the proof of the semi-conjugation property~(\ref{rel}).

\section{Conclusions}

  To summarize, we have shown that for the TASEP 
  with open boundaries the Markov matrices $M$ for  consecutive system sizes
 are related to each other  by a semi-conjugation  relation (\ref{rel}) via  two 
 different  transfer matrices  $T$ and $\tilde{T}$.
 An explicit form of $T$ is obtained from the matrix-product ansatz and the
 existence of the left transfer matrix $\tilde{T}$ is proved. 
 This relation between the dynamics corresponding to  two  consecutive system sizes
 is a consequence of the recursive structure of the Markov matrix. 
  We believe  that  this correspondance expresses a   fundamental property 
 which is at the heart of the exact  solution  of many nonequilibrium models. 
 One advantage of this approach is that  
  the existence of a  semi-conjugation expressed by  equation~(\ref{rel})
 can be investigated  on small system sizes by solving a set
  of linear equations, whereas we  do not know 
 how to test the existence of a matrix-product representation on small system sizes. 
 Above all,  we  were curious to understand the relation between the recursive structure
 of the  Markov matrix and the one implied by  the matrix-product representation. It seems
 to us that  the present work gives a partial  answer to this question. 
 Transfer matrices do  appear in other related models such
 as the multi-species ASEP on a ring \cite{sylvain}, 
 the  recently studied  annihilation model
 with exclusion  \cite{arvind} and for discrete-time
  dynamics such  as ordered sequential and fully parallel update 
 \cite{woelki-unpub}.  It would  also    be  of interest to extend this approach to 
 the partially asymmetric exclusion process with open boundaries.

\hfill\break

 We thank Arvind Ayyer for useful discussions and S. Mallick for a careful reading of the manuscript. Further MW likes to thank Prof. M. Schreckenberg for support.

\section*{Appendix} 

\hfill\break
{\it The matrix Ansatz for the TASEP:}

\hfill\break
 A configuration ${\mathcal C}$  of the TASEP on a discrete lattice
 of $L$ sites can be specified by assigning  the binary
 values of  $L$ local variables  $\tau_i$ with  $\tau_i =1$ if site $i$ is occupied
 and  $\tau_i =0$  otherwise.
 In the long time limit, the TASEP  reaches a steady state with
 a  non-trivial stationary measure:
  the  stationary
 probability $p({\mathcal C})$ of the  configuration  ${\mathcal C}$ 
  can be written as a matrix element over an ordered product of
$L$ matrices $E$ and $D$ representing empty and occupied sites
respectively:
\begin{eqnarray}
  p({\mathcal C}) = \frac{1}{Z_L} 
\langle W| \prod_{i=1}^L   \left( \tau_i D + (1-\tau_i) E \right)|V\rangle. 
\label{MatAns}
\end{eqnarray}
 where $Z_L$  is a normalisation factor. As shown in \cite{DEHP},
 the weights defined in~(\ref{MatAns})  correspond to the steady state
 probabilities if the 
 operators  $E$ and $D$ satisfy, along   with the left and right boundary
vectors $\langle W|$, $|V\rangle,$ the following  algebra
\begin{eqnarray}
\label{DE}
DE=D+E,\\
\label{WE}
\langle W|E=\alpha^{-1}\langle W|\\
\label{DV}
D|V\rangle =\beta^{-1}|V\rangle. 
\end{eqnarray}
 Thanks to  these   reduction relations, the weight of a configuration
 of size $L$  can be expressed as a linear combination
 of the  weights of  some configurations of size $L-1$. 
 This matrix-product representation yields 
 exact formulae for the currents, the density profiles,
  the steady correlations,  and allows to determine
 the exact phase diagram of the model \cite{DEHP}. 
 This algebra implies  the formula~(\ref{T}) for  the transfer matrix:
  the idea is to  always   apply  rule (\ref{WE})
   if possible;  with second priority apply (\ref{DE})
  and if this too  is   not possible then apply (\ref{DV}).

\hfill\break
{\it Proof that the  Kernel of $H_L$ is one dimensional:}

\hfill\break
 The matrix $H_L$, of size $2^L$, is defined in~(\ref{HL}).
 It can be considered   as the  `unevaluated determinant'
 of  the matrix $M_{L+1}$, given in equation~(\ref{M}). More precisely, we have:
\begin{equation}
 \det M_{L+1} = \det (- H_L) = \det  H_L  \, .
\end{equation}
  To write this equation we  treat  $M_{L+1}$ as a 2 by 2 block matrix
 whose elements are themselves matrices of size $2^L$ and  we apply 
 the following  theorem, proved in \cite{silvester}: if $A,B,C$ and  $D$ are square
 matrices  and  if  $C$ and $D$ commute, then 
\begin{equation}
\det \left( \begin{array} {cc} A  & B \\
C  &  D  \end{array}  \right) = \det \left( AD -BC  \right) \, .
\end{equation}
 Here $C= \alpha \thickone_{L}$. More generally, using the same theorem,
the characteristic polynomial  $\pi(X)$
 of $H_L$ can be written as
\begin{equation}
\pi(X) = \det (X\thickone_{L} - H_L) = \det \left[ M_{L+1} -\frac{X}{\alpha}
 \left( \begin{array} {cc} 0   & \thickone_{L}   \\ 0  &  0  \end{array}  \right)
 \right] \, .
\end{equation}
 The kernel of  $H_L$ is  of dimension 1 if $\pi(X)$ is divisible
 by $X$ but not by $X^2$. We show now that for small $X$,   $\pi(X)$ is of order
 $X$. We recall that   $\pi(X)$ is 
 given by the product of the eigenvalues of the matrix
 $ \left[  M_{L+1} -\frac{X}{\alpha}
 \left( \begin{array} {cc} 0   & \thickone_{L}   \\ 0  &  0  \end{array}  \right) \right] $. 
 For $X=0,$ we know that the spectrum of  $M_{L+1}$ consists of the eigenvalue 0
 (with multiplicity 1)  and that
 all the  other  eigenvalues have strictly negative real parts (Perron-Frobenius
 Theorem, see e.g. \cite{vankampen}). 
 For $X$ sufficiently small the non-zero eigenvalues will remain
 away from 0; the zero  eigenvalue  of  $M_{L+1}$ 
  will now be given by $E_0(X)$ 
 that can be calculated at  first order in $X$ using  perturbation theory:
 \begin{equation}
  E_0(X) = \frac{\langle 0| K |0 \rangle}{\langle 0| 0\rangle} + {\mathcal O}(X^2)
 = -\frac{X}{\alpha} \,\, 
 \langle 0| 
 \left( \begin{array} {cc} 0  & \thickone_{L}  \\ 0 & 0  \end{array}  \right)
 |0\rangle + {\mathcal O}(X^2)  \, ,
\end{equation}
 where $\langle 0| =(1,1,\ldots,1)$ and  $|0 \rangle$ 
 are the left and right eigenvectors 
 of  $M_{L+1}$  associated with  the eigenvalue 0 (i..e $|0 \rangle$  is the 
 steady state probability vector). Both these vectors have strictly positive
 entries and therefore  the dominant contribution to
 $E_0(X)$ is  of order  $X$. We have  shown that 
 the expansion of the characteristic polynomial  $\pi(X)$ for small $X$ 
 begins with a term of order $X$ with a non-vanishing coefficient: this proves that 0 is
  a simple eigenvalue of  $H_L$. 

\hfill\break
{\it Explicit construction of the Kernel of $H_L$:}

\hfill\break
 Let us call  $\left\{h_L(\tau_1, \dots, \tau_L)\right\}$ the components   of a non-zero
 vector  $h_L$ in the Kernel  of $H_L$. These numbers can be obtained  in a
 matrix-product form: 
\begin{equation}
 h_L(\tau_1, \dots, \tau_L)=
\langle \tilde{W}|\prod_{i=1}^{L}\left[\tau_i D+(1-\tau_i)E\right]|V\rangle \, ,
\end{equation}
 where $D,E$ and $|V\rangle$ are the same as in (\ref{DE}--\ref{DV}) and 
 $\langle \tilde{W}| = \langle {W}| D$. One has to show that  $H_L$
 acting on $h_L$  gives the null-vector. 
 The vector $h_L$ can be written as a column-vector
\begin{equation}
        h_L = \left(\begin{array} {l} \langle \tilde{W}|E {\mathcal S}_{L-1} |V\rangle\\
      \langle \tilde{W}|D{\mathcal S}_{L-1} |V\rangle  \end{array} \right)
\end{equation}
where ${\mathcal S}_{L-1}$ stands for all possible strings made of $L-1$ symbols $D$ and $E$.
Then, we have 
\begin{equation}
    M_{L} \left(\begin{array} {l} \langle  \tilde{W}|E {\mathcal S}_{L-1} |V\rangle\\
      \langle \tilde{W}|D{\mathcal S}_{L-1} |V\rangle  \end{array} \right)
 = \left(\begin{array} {l}  \langle \tilde{W}|{\mathcal S}_{L-1} |V\rangle -
  \alpha \langle  \tilde{W}|E {\mathcal S}_{L-1} |V\rangle\\
    \alpha \langle  \tilde{W}|E {\mathcal S}_{L-1} |V\rangle\ - 
      \langle \tilde{W}|{\mathcal S}_{L-1} |V\rangle  \end{array} \right) \, .
\label{actML}
\end{equation}
This equation is derived by considering  the  generic expression 
${\mathcal S}_{L-1}=D^{n_1}E^{n_2}\ldots E^{n_k}$  (with $n_1+\ldots+n_k =L-1$).
 We then  observe, as in \cite{DEvans}, that  through the action of  $M_{L}$    all terms
 in the bulk cancel out (thanks to the algebra  $DE = D+E$),  all terms
 on the right boundary cancel out (because of the rule~(\ref{DV}))  and only the
 left boundary terms do not simplify because the  bra-vector is 
 $\langle \tilde{W}|$  instead of  $\langle  {W}|$. It is important to remark
 that the precise form of  $\langle \tilde{W}|$ has not played any role
 in the derivation of equation~(\ref{actML}) and that this  relation 
 would remain  true if   $\langle \tilde{W}|$  were  replaced by an arbitrary bra-vector.

  From this result, we deduce, using equation~(\ref{def:nu-rho-omega}),  that
\begin{equation}
   \nu_L M_{L}\left(\begin{array} {l} \langle \tilde{W}|E {\mathcal S}_{L-1} |V\rangle\\
      \langle \tilde{W}|D{\mathcal S}_{L-1} |V\rangle  \end{array} \right) = 0 \, .
\label{actnuL}
\end{equation}
 Besides, we have
\begin{equation}
   ( M_{L} + \omega_L)
 \left(\begin{array} {l} \langle  \tilde{W}|E {\mathcal S}_{L-1} |V\rangle\\
      \langle  \tilde{W}|D{\mathcal S}_{L-1} |V\rangle  \end{array} \right) = 
  \left(\begin{array} {l}  \langle  \tilde{W}|{\mathcal S}_{L-1} |V\rangle -
   \langle  \tilde{W}|E {\mathcal S}_{L-1} |V\rangle  \\
    -    \langle  \tilde{W}|{\mathcal S}_{L-1} |V\rangle  \end{array} \right) 
 = -  \left(\begin{array} {l} \langle  {W}|E {\mathcal S}_{L-1} |V\rangle\\
      \langle  {W}|D{\mathcal S}_{L-1} |V\rangle  \end{array} \right) \, ,
\label{actMplusOmega}
\end{equation}
where we have used  $ \langle  \tilde{W}| =  \langle  {W}|D$
 and $DE = D+E$  in the last equality.
 Combining equations~(\ref{actnuL}) and ~(\ref{actMplusOmega}),
 and using equation~(\ref{HL}),  we conclude that
\begin{equation}
H_L  \left(\begin{array} {l} \langle  \tilde{W}|E {\mathcal S}_{L-1}  |V\rangle\\
      \langle  \tilde{W}|D{\mathcal S}_{L-1} |V\rangle  \end{array} \right) =
  \left( \nu_{L} M_{L} - M_{L} (M_{L}+\omega_{L} )\right)
\left(\begin{array} {l} \langle \tilde{W}|E {\mathcal S}_{L-1} |V\rangle\\
      \langle  \tilde{W}|D{\mathcal S}_{L-1} |V\rangle  \end{array} \right) =
M_{L} \left(\begin{array} {l} \langle  {W}|E {\mathcal S}_{L-1} |V\rangle\\
      \langle  {W}|D{\mathcal S}_{L-1} |V\rangle  \end{array} \right) = 0\, .
\end{equation}
The last equality is true because the vector on which $M_{L}$ acts is  precisely
 the TASEP steady state vector,  given 
 by the  algebra~(\ref{DE}--\ref{DV}).

\hfill\break

\hfill\break
{\it Proof that the  Kernel of $H_L$ is contained in the Kernel of $G_L$:}

\hfill\break
 Using the recursive formula~(\ref{recursGL}) we can write  $G_L$
 as follows:
\begin{eqnarray}
G_{L}=  \kappa_L M_L + \left(\begin{array} {ll}
 0  & 0\\
 0 & G_{L-1}\end{array}\right)  \quad \hbox{ with } \quad
 \kappa_L =  \left(\begin{array} {ll}
 \thickone_{L-1}  & \thickone_{L-1}\\
 0 & 0\end{array}\right) \, .
\label{recurs2GL}
\end{eqnarray}
 From equation~(\ref{actML}),  we deduce that  $(\kappa_L M_L)h_L =0$ and this 
 would remain  true if $\langle  \tilde{W}|$ were  replaced by  any bra-vector.
The action of $G_{L}$ on  $h_L$ is  thus given by
\begin{eqnarray}
  G_{L} \left(\begin{array} {l} \langle \tilde{W}|E {\mathcal S}_{L-1} |V\rangle\\
      \langle  \tilde{W}|D{\mathcal S}_{L-1}  |V\rangle  \end{array} \right)
 = \left(\begin{array} {ll} 0  & 0\\  0 & G_{L-1}\end{array}\right)
 \left(\begin{array} {l} \langle  \tilde{W}|E {\mathcal S}_{L-1} |V\rangle\\
      \langle  \tilde{W}|D{\mathcal S}_{L-1} |V\rangle  \end{array} \right)
=  \left(\begin{array} {l}  0 \\
     G_{L-1}  \langle  \tilde{W}|D{\mathcal S}_{L-1} |V\rangle  \end{array} \right) \,.
\label{actionrecGL}
\end{eqnarray}
 This identity is satisfied regardless of the
 precise form of the  bra-vector $\langle  \tilde{W}|$.
  Finally,  rewriting  the string
 ${\mathcal S}_{L-1} =  
\left(\begin{array} {l} E {\mathcal S}_{L-2} \\
      D{\mathcal S}_{L-2}  \end{array} \right)$ and defining 
 $\langle   \tilde{\tilde{W}}| = \langle  \tilde{W}| D $,
 the last equation can be recasted  in a form that makes its 
 recursive structure clear
\begin{eqnarray}
 G_{L} h_L =  \left(\begin{array} {l} \quad \quad \quad \quad  0 \\
     G_{L-1}  \left(\begin{array} {l} 
 \langle  \tilde{\tilde{W}}| E {\mathcal S}_{L-2}  |V\rangle \\
    \langle  \tilde{\tilde{W}}| D {\mathcal S}_{L-2}  |V\rangle     \end{array} \right)
 \end{array} \right) \,. 
\end{eqnarray}
 Using iteratively equation~(\ref{actionrecGL}), we conclude that  $G_{L} h_L = 0$,
 i.e. that the kernel of $H_L$ is included in that of  $G_L$ and that, therefore,
 the equation~(\ref{GRH}) is solvable.



\begin{thebibliography}{10}
\bibitem{spohn} Spohn H 1991 {\it Large Scale Dynamics of Interacting Particles}
 (Berlin: Springer).
\bibitem{liggett}   Liggett T M  1999  {\it Stochastic Interacting Systems:
 Contact, Voter and Exclusion  Processes}  (Berlin: Springer).
\bibitem{PaulK} Krapivsky P, Redner S and Ben-Naim E  2010 
  {\it Kinetic View of Statistical Physics}
  (Cambridge: Cambridge University Press)  {\it (to appear)}.
\bibitem{zia} Schmittmann B and Zia R K P 1995 
  {\it  Statistical Mechanics of Driven Diffusive Systems
 (Phase Transitions and Critical Phenomena vol 17)} (London: Academic Press Inc.).
\bibitem{derrida} Derrida B 1998 {\it Phys. Rep.} {\bf 301}  65.
  \bibitem{EvansBraz}  Evans M R 2000 {\it Braz. J. Phys.} {\bf 30} 42.
\bibitem{SchutzRev} Sch\"utz G M  2000  {\it Exactly Solvable Models
 for Many-Body Systems Far From  Equilibrium (Phase Transitions and
  Critical Phenomena vol 19)} (London: Academic Press Inc.).
\bibitem{varadhan} Varadhan S R S 1996,   {\it The complex
 story of simple exclusion} in {\it Ito's stochastic calculus},
 Fukushima et al. eds.  (Berlin: Springer).

\bibitem{Dhar} Dhar D  1987      {\it Phase Transitions} {\bf 9} 51.
\bibitem{GwaSpohn} Gwa L H  and Spohn H 1992  {\it Phys. Rev. A} {\bf 46} 844.
\bibitem{DEHP} Derrida B,  Evans M R, Hakim V and Pasquier V 1993
 {\it J. Phys. A} {\bf 26} 1493.
\bibitem{DDM} Derrida B, Domany E and Mukamel D 1992 {\it J. Stat. Phys.} {\bf 69} 667.
\bibitem{Schutz} Sch\"utz G M  and  Domany E 1993  {\it J. Stat. Phys.} {\bf 72} 277.
\bibitem{Blythe} Blythe R A and Evans M R 2007 {\it J Phys. A} {\bf 40} R333.
\bibitem{Derrida2class} Derrida B, Janowsky S A, Lebowitz J L and Speer E R 1993
 {\it J. Stat. Phys.} {\bf 73} 813.
\bibitem{degier} de Gier J and Nienhuis B 1999 {\it Phys. Rev. E} {\bf 59} 4899.
\bibitem{MMR} Mallick K, Mallick S and  Rajewsky N 1999  {\it J. Phys. A}  {\bf 32} 8399. 
\bibitem{chikashi1}  Arita C 2006 {\it J. Stat.  Mech.:Theor. Exp} P12008. 
\bibitem{woelkirand} Woelki M and Schreckenberg M 2009 {\it J. Phys. A} {\bf 42} 325001.
\bibitem{woelkipar} Woelki M and Schreckenberg M 2009 {\it J. Stat. Mech.} P05014 
\bibitem{hinrichsen} Hinrichsen H, Sandow S and Peshel I 1996
  {\it J. Phys. A}  {\bf 29} 2643.
\bibitem{nikolaus} Rajewsky N, Santen L, Schreckenberg M and Schadschneider A 1998 
{\it J. Stat. Phys.} {\bf 92} 151.
\bibitem{krebs} Krebs K and Sandow S 1997 {\it J. Phys. A} {\bf 30} 3165. 
\bibitem{pablo}   Evans M R, Ferrari P A  and  Mallick K 2009 
 {\it J. Stat. Phys.} {\bf 135} 217.
\bibitem{sylvain}  Prolhac S,  Evans M R  and  Mallick K  2009 {\it J. Phys. A}
{\bf 42} 165004.
\bibitem{chikashi}  Arita C,  Kuniba A,  Sakai K and  Sawabe T 2009 
 {\it J. Phys. A} {\bf 42}  345002.
\bibitem{arvind}  Ayyer A and Mallick K 2009 {\it J. Phys. A}  (to appear) arXiv
0910.0693v1

 \bibitem{silvester} Silvester J R 2000 {\it Math. Gazette} {\bf 84} 460. 
  \bibitem{vankampen}  Van Kampen N G  1992  {\it Stochastic Processes in
  Physics and Chemistry} (Amsterdam: North-Holland).
\bibitem{woelki-unpub} Woelki M {\it unpublished}
 \bibitem{DEvans} Derrida B and  Evans M R  1996 in {\it Nonequilibrium
  Statistical Mechanics in One  Dimension}  (Cambridge: Cambridge University Press).



\end{thebibliography}
\end{document}